\documentclass[a4paper,10pt]{article}
\usepackage[dvips]{graphicx}
\newcommand{\be}{\begin{equation}}
\newcommand{\ee}{\end{equation}}
\newcommand{\bey}{\begin{eqnarray}}
\newcommand{\eey}{\end{eqnarray}}
\newcommand{\bi}{\bibitem}
\usepackage{amsmath}
\usepackage{amsfonts}
\usepackage{graphics}
\usepackage{amssymb}
\usepackage{textcomp}
\title{Why Noether symmetry of F(R) theory yields three-half\\
 power law?}
\author{ Kaushik Sarkar$^{\dag}$, Nayem Sk$^{\ddag}$, Soumendranath Ruz$^{\S}$, Subhra Debnath$^{*}$\\
and Abhik Kumar Sanyal$^{\star}$ }

\begin{document}
\maketitle
\begin{center}

$\dag, \ddag, \S$Dept. of Physics, University of Kalyani, Nadia, India - 741235.\\
$*,\star$Dept. of Physics, Jangipur College, Murshidabad, India - 742213.\\
\end{center}

\footnote{
Electronic address:\\
\noindent $^{\dag}$sarkarkaushik.rng@gmail.com\\
\noindent $^{\ddag}$nayemsk1981@gmail.com\\
\noindent $^{\S}$ruzfromju@gmail.com\\
\noindent $^{\star}$subhra\_ dbnth@yahoo.com\\
\noindent $^{*}$sanyal\_ ak@yahoo.com}\\

\begin{abstract}

\noindent
Noether symmetry of $F(R)$ theory of gravity in vacuum or in matter dominated era yields $F(R) \propto R^{\frac{3}{2}}$. We show that this particular curvature invariant term is very special in the context of isotropic and homogeneous cosmological model as it makes the first fundamental form $h_{ij}$ cyclic. As a result, it allows a unique power law solution, typical for this particular fourth order theory of gravity, both in the vacuum and in the matter dominated era. This power law solution has been found to be quite good to explain the early stage but not so special and useful to explain the late stage of cosmological evolution. The usefulness of Palatini variational technique in this regard has also been discussed.
\end{abstract}

\section{Introduction}
In the last decade several path finder models have been explored which enunciated the best fit with SNIa data set. For most of these models  SNIa data set implies a smooth transition from a decelerated phase to a recent accelerated phase of cosmic evolution, and the presence of nearly $74\%$ of dark energy, in disguise. However, starting from the $\Lambda CDM$ concordance model, often cited as the standard model, all the models suffer from some sort of disease. The order of magnitudes of the vacuum energy density (corresponding to cosmological constant $\Lambda$) calculated by the Quantum field theorists and the Cosmologists differ by a huge ($10^{120}$) amount. Quintessence model solves the problem, but it requires a scalar field with some exotic potential, which is not at hand presently. Further, Quintessence model does not suffice to handle the situation with state parameter $\omega < -1$, which is not excluded in view of the presently available data. It requires additional scalar field containing kinetic term with reverse sign. Thus, the situation gets more and more complicated. There was attempt to solve the puzzle assuming particle creation phenomena at the expense of gravitational field \cite{lima2006,subhra2011} which has strong theoretical basis, but there is no direct experimental evidence as yet. The only thing that is left is curvature, which can not be detected directly. Inflation in the early Universe may be an artefact of $R^2$ term in gravity \cite{staro1980} and comic acceleration in the late Universe might have been caused by $R^{-1}$ term \cite{carroll2004}. Thus, it appears that an action in the form $S = \int[\alpha R + \beta R^2 + \gamma R^{-1}]\sqrt{-g} d^4 x$, having pure gravitational origin, might be the one to solve the cosmic puzzle from early Universe till date.\\
However, $R^{-1}$ term fails to produce Newtonian gravity in the weak energy limit and does not comply with the solar test \cite{cj2003}. Even worse is, it shows unavoidable instabilities within matter in the weak gravity limit \cite{dol2003} and fails to explain big bang nucleosynthesis (BBN) \cite{ab2006}. Further, for any $R^{-n}$ term, with $n > 0$, it has been shown \cite{amen2007} that the scale factor during the matter phase grows as $t^{\frac{1}{2}}$, instead of the standard $t^\frac{2}{3}$ - law, prior to the transition to an accelerated expansion. This is grossly inconsistent with cosmological observations viz., WMAP (Wilkinson Microwave Anisotropy Probe) and LSS (Large Scale Structure). Therefore it is required to find some scalar invariant other than $R^{-1}$, suitable to explain late time cosmic acceleration and should have Newtonian analogue in the weak field limit to pass the solar test. Such a theory, dubbed as $F(R)$ theory of gravity \cite{sotiriou2007, nojiri2011} is much more attractive, promising and intriguing than scalar field models. But then, there are at least a couple of associated problems. Firstly, out of indefinitely many curvature invariant terms, one requires an elegant method to choose suitable ones that fit all cosmological data at hand. (Remember that in the early Universe, higher order terms were incorporated to obtain a renormalizable theory of gravity). Next, it is practically impossible to solve the field equations of $F(R)$ theory of gravity, corresponding to any nonlinear form of $R$. Both the problems presumably may be solved by invoking Noether symmetry. A review in this connection has recently been presented by Capozziello et al \cite{cap2012}. \\
In order to express higher order theory of gravity in non-degenerate and canonical form, it is required to be spanned by an auxiliary variable in addition to the basic variables. In the process, the configuration space of the Lagrangian gets enlarged. Thus, one can demand the existence of Noether symmetry amongst the field variables (viz., between the scale factor and the auxiliary variable) even when the theory is reduced to a measure zero subset of its configuration space, i.e., if the Robertson-Walker minisuperspace model is accounted for. This was attempted largely by several authors \cite{caplam2000, capfel2008, vak2008a, vak2008b, cap2008} to find a suitable form of $F(R)$, spanning the Lagrangian by a set of configuration space variables $(a, \dot a, R, \dot R)$, where $R$ acts as the auxiliary variable. Noether symmetry of $F(R)$ theory of gravity \cite{caplam2000, capfel2008, vak2008a, vak2008b, cap2008} yields $F(R) \propto R^{\frac{3}{2}}$ in the Robertson-Walker minisuperspace model, which, as we shall show, turns out to be important in studying early Universe but not the late. Although, explicit solution indicates a smooth transition from early deceleration to the late time cosmic acceleration \cite{cap2008}, which is definitely an attractive feature of such a curvature invariant term, nevertheless, the more interesting issue is that, out of infinite number of curvature invariant terms, Noether symmetry selects $R^{\frac{3}{2}}$ term in particular. Thus, we think that the study of Noether symmetry of $F(R)$ theory of gravity requires further attention.\\
Despite the fact that $R^{\frac{3}{2}}$ theory of gravity has some attractive features, particularly in explaining the late time cosmic evolution, nevertheless, here we observe that it suffers from the same type of disease associated with $R^{-n}$ theories of gravity ($n > 0$), which is also apparent from the solution already obtained by \cite{cap2008}. The solution depicts that in the matter dominated era, prior to the transition to an accelerated expansion, the scale factor never follows the standard $t^{\frac{2}{3}}$-law. This disease may be cured if such an action would have been supplemented by Einstein-Hilbert term. Hence, in the following section 2, we try to explore Noether symmetry of $F(R)$ gravity being supplemented by the Einstein-Hilbert term. This is done to understand if the technique of finding such symmetry is correct. If it is, then the same old result is expected, since $\alpha R + \beta F(R) = \gamma F_1(R)$ and this is what we have obtained. For a further check, in section 3, we explore Noether symmetry of $R^{\frac{3}{2}}$ term following the standard technique and end up with the same solutions obtained earlier \cite{cap2008}. Thus we understand that Noether symmetry of pure gravity yields nothing more than $F(R) = R^{\frac{3}{2}}$. This initiates to explore the very speciality of this particular curvature invariant term. In section 4, under a change of variable $h_{ij} = z = a^2$, we find that Noether symmetry is in-built in $R^{\frac{3}{2}}$ term, since $h_{ij} =z = a^2$ becomes cyclic. In the process, the field equations are solved quite easily and the same solution emerges. This implies that one does not require to request \cite{cap2008} Noether symmetry of $R^{\frac{3}{2}}$ curvature invariant term. In section 5, we review Noether symmetry of $F(R)$ theory of gravity, in view of the changed variable $h_{ij}$ and observe that the conserved current may be solved quite easily and for the purpose, it is not required to find cyclic variable any more. In the next section 6, we introduce matter and find solutions both in the radiation and in the matter dominated era. Section 7 is devoted to explore the viability of $R^{\frac{3}{2}}$ term in the context of presently available cosmological data. In section 8, we discuss the usefulness of the result obtained invoking Noether symmetry in the Palatini variational technique. We conclude with our findings in section 9. The most important finding is that, this particular curvature invariant term, which leads to fourth order gravity, yields only a unique solution both in vacuum and in the matter dominated era and unless the action is supplemented by a linear term, it does not produce a viable cosmological model.

\section{In search of Noether symmetry for Einstein-Hilbert action being modified by $F(R)$ theory of gravity.}

We have already discussed the problem of $F(R)$ theory of gravity if it does not contain Einstein-Hilbert part in the action. Noether symmetry of $F(R)$ theory of gravity yields $F(R) = R^{\frac{3}{2}}$. The solution is encouraging as far as late time cosmic acceleration is concerned. But the problem is that, it only passes transiently through $a \propto t^{\frac{2}{3}}$ in the matter era before the transition to accelerated phase \cite{cap2008} and thus fails to explain WMAP and LSS, as in the case of $R^{-n}$ term \cite{ab2006,amen2007}. So, we try to find Noether symmetry for the $F(R)$  theory being coupled to the Einstein-Hilbert sector. Although it appears to be trivial, since these two terms together may be combined to get yet another $F(R)$ theory. Nevertheless, it turns out to be an important test to check if the technique of exploring Noether symmetry is correct. Hence we start with the following action,
\be A_1=\int\sqrt{-g}\;d^4 x\Big[\frac{R}{16\pi G}+ B F(R)\Big] - \frac{1}{8\pi G}\int [\sqrt h K]d^3 x - 2 B\int \Big[\sqrt h F_{,R} K \Big]d^3 x,\ee
where, $\frac{1}{8\pi G}\int [\sqrt h K]d^3 x$ is the Gibbons-Hawking boundary term and $\Sigma = 2 B\int \left[\sqrt h F_{,R} K \right]d^3 x,$ is the surface term for $F(R)$ theory of gravity that emerges under variational principle (metric formalism). In the above, $h$ is the determinant of the metric on three space, $K$ is the trace of the extrinsic curvature and $F_{,R}$ is the derivative of $F(R)$ with respect to $R$. Note that this surface term $\Sigma$ reduces to the Gibbons-Hawking boundary term taking $F(R) = \frac{1}{16\pi G} R$ and that for the curvature squared action $F(R) = R^2$ \cite{aksbm2001,sanmod2002,aks2002,san2003,as2005,ak2011}. Action (1) in the background of isotropic and homogeneous Robertson-Walker line element,
\be dS^2 = -dt^2+a^2\left[\frac{dr^2}{1-kr^2}+r^2 d\theta^2 + r^2\sin^2\theta d\phi^2\right],\ee
leads to
\[ A =\int\Big[\frac{6}{16\pi G}\Big(\frac{\ddot a}{a}+\frac{\dot a^2}{a^2}+\frac{k}{a^2}\Big)+BF(R)-\lambda\Big\{R - 6\Big(\frac{\ddot a}{a}+\frac{\dot a^2}{a^2}+\frac{k}{a^2}\Big)\Big\}\Big]a^3 dt \]\be - \frac{1}{8\pi G}\int \sqrt h Kd^3 x- 2B \int \sqrt h F_{,R} K d^3 x,\ee
where, the expression for $R$ has been introduced as a constraint through Lagrange multiplier $\lambda$. Varying the above action with respect to $R$, one gets $\lambda = BF,_R$, which is substituted in the action. Now under integration by parts the surface terms get cancelled and the action may be expressed as (taking $\frac{3}{8\pi G} = M$),
\be A =\int\Big[-Ma\dot a^2+ Mka+Ba^3 (F - R F,_R) - 6B a \dot a^2 F,_R - 6B a^2\dot a F,_{RR} \dot R+6B k a F,_R \Big] dt\ee
while the Noether equation is,
\[X L =\pounds_X L =  \alpha \Big(-M\dot a^2+ Mk + 3Ba^2 (F - R F,_R)- 6B  \dot a^2 F,_R-12 B a\dot a \dot RF,_{RR} +6B k F,_R\Big) \]
\[ + \beta B\Big((6ka - a^3 R - 6 a \dot a^2) F,_{RR} - 6 a^2\dot a \dot R F,_{RRR} \Big) + \Big(\frac{\partial\alpha}{\partial a}\dot a+\frac{\partial\alpha}{\partial R}\dot R \Big)(-2Ma\dot a-12Ba\dot aF,_R
 -6Ba^2 \dot RF,_{RR}) \]\be+ \Big(\frac{\partial\beta}{\partial a}\dot a+\frac{\partial\beta}{\partial R}\dot R \Big)(-6B a^2\dot a F,_{RR})=0,\ee
where, $X$ is the vector field given by
\[X = \alpha(a,R)\frac{\partial}{\partial a} + \beta(a,R)\frac{\partial}{\partial R}+\dot\alpha(a,R)\frac{\partial}{\partial \dot a} + \dot\beta(a,R)\frac{\partial}{\partial \dot R}\;.\]
Now we equate the coefficients of $\dot a^2,\;\dot R^2,\; \dot a\dot R$ along with the term free from derivative respectively to zero as usual to obtain,
\be (M + 6BF,_R )(\alpha + 2a \alpha ,_a)+ 6B(\beta  + a \beta ,_{a})a F,_{RR}  = 0. \ee
\be 6Ba^2F,_{RR}\alpha ,_{R} = 0. \ee
\be  6 B \beta a^2F,_{RRR} +6B(2\alpha   +  a \alpha ,_{a} + a \beta,_R)a F,_{RR}+ 2(M + 6 B  F,_R)a \alpha ,_R = 0. \ee
\be\alpha [M k + 3 B(F - R F,_R)a^2 + 6BkF,_R] + B\beta [6k-a^2 R] a F,_{RR} = 0. \ee
Equation (7) implies $\alpha \ne \alpha (R)$, since, we require nonlinear form of $F(R)$, ie., $F,_{RR}  \ne 0$. So, we have, $\alpha = \alpha(a)$, and separating $\beta$ as $\beta = A(a) S(R)$. Equation (8) is expressed as,
\be S\frac{F,_{RRR}}{F,_{RR}}+\frac{d S}{d R}=-\frac{1}{A}\left[2\frac{\alpha}{a}+\frac{d\alpha}{da}\right] = c_1,\ee
where, $c_1$ is a separation constant. The above equation (10) yields,
\be \frac{d\alpha}{da}+2\frac{\alpha}{a} = -c_1A,\ee
and
\be \frac{F,_{RRR}}{F,_{RR}}+\frac{S,_R}{S} = \frac{c_1}{S}.\ee
Equation (6) can as well be separated using the separation constant $c_2$ as,
\be S = \frac{\frac{M}{6B}+F,_R}{c_2F,_{RR}},\ee
and
\be a^2A,_a+aA + c_2(2a\alpha,_a+\alpha)=0.\ee
Equations (11) and (14) are combined to give the following differential equation,
\be a^2\alpha,_{aa}+(3-2c_1c_2)a\alpha,_a-c_1c_2\alpha=0\ee
In view of equations (11) and (13), equation (9) takes the following form,
\be k(M+6BF,_R)[(c_1c_2-2)\alpha-a\alpha,_a)] + 3Bc_1c_2(F-R F,_R)\alpha a^2+\frac{(M+6BF,_R)}{6}(a\alpha,_a+2\alpha)a^2R = 0.\ee
Thus, one can solve $\alpha$ in view of equation (15) and hence $A(a)$ from (11). $F(R)$ and $S(R)$ may be solved in view of equations (12) and (13). Finally, equation (16) is satisfied for $c_1 c_2 = 2$. At the end, following set of solutions exists for $k = \pm 1, 0$.
\be\alpha = \frac{n}{a},\;\;\;\;\beta = -\frac{2n R}{a^2},\;\;\;\; {\mathcal F} = \frac{d}{dt}(a\sqrt R),\ee
where, $n$ is a constant of integration and ${\mathcal F}$ is the conserved current. However, the most interesting result is,
\be F(R) = -\frac{M}{6B} R + C R^{\frac{3}{2}},\ee
which, under the substitution $M = \frac{3}{8\pi G}$, reduces the action to,
\be A = \int \sqrt{-g}~ d^4 x[D R^{\frac{3}{2}}].\ee
Thus, Noether symmetry of $F(R)$ does not admit an explicit Einstein-Hilbert term in the $F(R)$ theory of gravity. This result is expected and it proves that the technique of exploring Noether symmetry is correct. So, in the following sections we concentrate upon our main concern of exploring the very speciality of $R^{\frac{3}{2}}$ term.

\section{Noether symmetry of $R^{\frac{3}{2}}$ - standard technique.}

This section is important for the following reason. To express $F(R)$ theory of gravity in non-degenerate and canonical form, one usually introduces the form of $R$ as a constraint through a Lagrange multiplier $\lambda$. The extended configuration space in such case is $(a,~\dot a,~R,~\dot R)$. In the standard formalism, where a form of curvature invariant term is fixed, the configuration space is different viz., $(a,~\dot a,~q,~\dot q)$. Here, $q$ is an auxiliary variable chosen as the derivative of the action with respect to the highest derivative of the variable appearing in the action, that has not been possible to integrate by parts to produce a surface term. Here we check, if Noether symmetry along with the solutions presented by \cite{cap2008} exists, when action containing $R^{\frac{3}{2}}$ term is spanned by such a set of configuration space variables, i.e., whether Noether symmetry is an artifact of a particular choice of configuration space variables or not. One may expect that since Noether symmetry of $F(R)$ theory yields $R^{\frac{3}{2}}$, so $\pounds_X L = 0$ will be exactly solved if $F(R) \propto R^{\frac{3}{2}}$ is assumed a-priori. The action under consideration in the Robertson-Walker minisuperspace model is,
\be A = B\int  R^{\frac{3}{2}}\sqrt{-g}d^4 x + 2B \int \left[\sqrt h F_{,R} K \right]d^3 x = B\int 6^{\frac{3}{2}}[a\ddot a+\dot a^2+k]^{\frac{3}{2}}dt - 9\sqrt 6 B[a\ddot a+\dot a^2+k]^{\frac{1}{2}}a\dot a.\ee
Now let us define the auxiliary variable $Q$ as,
\be q = \frac{\partial S}{\partial \ddot a} = 9\sqrt 6 B[a\ddot a+\dot a^2+k]^{\frac{1}{2}}a = 9B a^2\sqrt R,\ee
which is radically different from $R$. Expressing the action in non-degenerate, canonical form in view of the above auxiliary variable
\be A = \int\left[\frac{q}{a}[a\ddot a+\dot a^2+k]-\frac{1}{1458 B^2}\frac{q^3}{a^3}\right]dt - 9\sqrt 6 B[a\ddot a+\dot a^2+k]^{\frac{1}{2}}a\dot a\ee
and integrating by parts, we obtain,
\be A = \int\left[-\dot q\dot a + \frac{q}{a}(\dot a^2 + k)-\frac{1}{1458 B^2}\frac{q^3}{a^3}\right]dt + q\dot a  - 9\sqrt 6 B[a\ddot a+\dot a^2+k]^{\frac{1}{2}}a\dot a.\ee
One can easily check that the last two terms get cancelled and so, there is no controversy on the choice of the auxiliary variable. The field equations are,
\be \ddot q - 2\dot q\frac{\dot a}{a} - 2q\frac{\ddot a}{a} + q\frac{\dot a^2}{a^2} - q\frac{k}{a^2} + \frac{1}{486 B^2}\frac{q^3}{a^4} = 0.\ee
\be-\dot a\dot q +  q\frac{\dot a^2}{a} - q\frac{k}{a} + \frac{1}{1458 B^2}\frac{q^3}{a^3} = 0.\ee
Equation (24) is a fourth order equation in the scale factor ($a$), while equation (25) is third order and apparently it is impossible to solve the pair. So, it is better to ask Noether symmetry between the field variables $a$ and $q$, i.e., $\pounds_X L = XL = 0$, which reads,
\be\alpha\left[\frac{q^3}{486 B^2 a^4}-\frac{\dot a^2 + k}{a^2}q \right]+\beta\left[\frac{\dot a^2+k}{a} - \frac{q^2}{486 B^2 a^3}\right]+\left(\alpha_a\dot a+\alpha_q\dot q\right)\left[2\frac{\dot a}{a}q-\dot q\right]-\left(\beta_a\dot a+\beta_q\dot q\right)\dot a =0.\ee
Equating coefficients of $\dot a^2$, $\dot q^2$, $\dot a\dot q$ and terms free from derivative, we get following four equations, which are linear in $\alpha$ and $\beta$, viz.,
\be \alpha q - 2 a\alpha_{,a} q - a\beta + a^2\beta_{,a}=0.\ee
\be \alpha_{,q} =0.\ee
\be \alpha_{,a} -2\frac{q}{a}\alpha_{,q} +\beta_{,q} = 0.\ee
\be (\alpha q-\beta a)\left(486 B^2 k a^2 - q^2\right) =0.\ee
In the above comma represents derivative, as before. Equation (28) dictates that $\alpha \ne \alpha(q)$, while equation (30) clearly yields two cases, viz.,
\[\beta = \frac{\alpha}{a}q,\;\;\;\;and,\;\;\;\;q = 9\sqrt 6 B\sqrt{k}~a.\]

\subsection{\textbf{Case - I}, $\beta = \frac{\alpha}{a}q$}

Equation (29) in view of the above condition yields,,

\be \alpha = \frac{\alpha_{0}}{a},\;\;\; and~ so,\;\;\;\beta = \frac{\alpha_{0}}{a^2}q.\ee
It is quite easy to verify that equation (27) is trivially satisfied for the above solutions. Thus the conserved current $\mathcal F$ is,

\be  {\mathcal F} = \alpha p_{a} + \beta p_{q}  \Longrightarrow \frac{d}{dt}\left(\frac{q}{a}\right) = Constant \Rightarrow \frac{d}{dt}(a\sqrt R) = Constant,\ee
where, the final form of the conserved current, which is the same old one found in (17), is obtained by plugging in the definition of $q$ from relation (21). One can immediately solve the above equation for $a$ as,

\be a = \left[a_4 t^4+a_3 t^3 + \left(\frac{3a_3^2}{8a_4} - k\right) t^2 + a_2 t + a_1\right]^{\frac{1}{2}},\ee
where, $a_1$, $a_2$, $a_3$ and $a_4$ are integration constants. It is not difficult to check that the above solution satisfies both the field equations (24) and (25). This is exactly the solution obtained and analyzed by \cite{cap2008}. Note that, the solution has been found directly from Noether symmetry, like the one obtained earlier in the case of $R^2$ theory of gravity being coupled with a scalar field \cite{aec2005} and the one with Gauss-Bonnet dilatonic interaction \cite{gbd2011} and we don't require to solve the field equations any more. This is a unique feature of higher order theory of gravity. The above vacuum solution may be treated as one suitable to explain the early Universe, when curvature played the dominant role. The Universe started evolving as $a \propto \sqrt t$ and soon transits to a power law inflationary era.

\subsection{\textbf{Case - II}, $q = 9\sqrt{6 k} B a$}

Since here $q$ is proportional to $a$, so, the field variables practically reduce to one, and it is nonsense to try to find Noether symmetry. Rather, we can check, if such a relation between $q$ and $a$ yields a solution in view of the definition of $q$ given in (21) that satisfies the field equations (24) and (25). The solution is,

\be a = \sqrt{d_1 t + d_0},\ee
where, $d_1$ and $d_0$ are integration constants. This solution satisfies the field equation (24), but equation (25) is satisfied only under the condition, $k = 0$. This is trivial, since in this case $q = 0$ and the field equations are trivially satisfied. In view of the definition of $q$, the Ricci scalar $R = 0$ and so, it automatically leads to Friedmann solution in the radiation dominated era in the following manner. The combination of Friedmann equations,
\[2\frac{\ddot a}{a} + \frac{\dot a^2}{a^2} + \frac{k}{a^2} = -8\pi G p\;\;and\;\;
3\left(\frac{\dot a^2}{a^2} + \frac{k}{a^2}\right)=  8\pi G \rho\]
together yield,
\[6\left(\frac{\ddot a}{a} + \frac{\dot a^2}{a^2} + \frac{k}{a^2}\right) = R = 8\pi G(\rho - 3p).\]
Now, in the radiation dominated era, $\rho - 3p = 0$, and so the above equation reduces to,
\[ 6\left(\frac{\ddot a}{a} + \frac{\dot a^2}{a^2} + \frac{k}{a^2}\right) = R = 0.\]
Further if $k = 0$, the solution of the above equation is $a \propto t^{\frac{1}{2}}$.

\section{What is so special in $R^{\frac{3}{2}}$? A different look under change of variable.}

In general, Noether symmetry may not exists in arbitrary configuration space variables, particularly if the change of variable is not canonical. However, in this case we observe that the same symmetry and the conserved current is found treating both $R$ and $q$ as auxiliary variables separately. Hence, the question that naturally arises, what is so special in the curvature invariant term $R^{\frac{3}{2}}$. In this section, we pose to answer this question. It has already been emphasized that the basic variable is the first fundamental form $h_{ij}$ and not the scale factor $a$  \cite{aksbm2001,sanmod2002,aks2002,san2003,as2005,ak2011}. So, here we start with the basic variable $h_{ij} = a^2 = z$. In view of the expression of Ricci scalar,
\[R = 6\left(\frac{\ddot z}{2z} + \frac{k}{z}\right).\]
The action now reads,
\be A = B\int 3\sqrt3(\ddot z + 2k)^{\frac{3}{2}}dt - 2 B\int \left[\sqrt h F_{,R} K \right]d^3 x.\ee
The auxiliary variable is
\be Q = \frac{\partial S}{\partial \ddot z} = \frac{9\sqrt 3}{2}B(\ddot z + 2k)^{\frac{1}{2}} = \frac{9}{2}B a\sqrt R.\ee
Note that $Q$ is again different from $R$ and $q$ used earlier. The canonical form of the action is
\be A = \int \Big[Q(\ddot z + 2k) - \frac{4Q^3}{729 B^2}\Big]dt - 2 B\int \left[\sqrt h F_{,R} K \right]d^3 x = \int\left[-\dot Q\dot z + 2k Q - \frac{4Q^3}{729 B^2}\right]dt.\ee
The definition of the auxiliary variable is restored immediately from the $Q$ variation equation, while the variable $z$ turns out to be cyclic. Thus the field equations are oversimplified to,
\be \ddot Q = 0 \Rightarrow \frac{d}{dt}(a\sqrt R) = Constant,\ee
\be \dot z\dot Q + 2kQ - \frac{4 Q^3}{729 B^2} = 0,\ee
where, equation (36) has been used and equation (38) turns out to be the same conserved current obtained earlier in equations (17) and (32). Equation (38) immediately gives the following solution,
\be a = \left[a_4 t^4 + a_3 t^3 + \left(\frac{3a_3^2}{8a_4}-k\right) t^2 + a_2 t + a_1\right]^{\frac{1}{2}},\ee
which resembles exactly with the solution (33) obtained earlier in view of Noether symmetry. Note that, equation (39) also
yields the same above solution. Thus, it is clear that Noether symmetry is in built in $R^{\frac{3}{2}}$ theory of gravity, since $z$ turns out to be cyclic automatically.

\section{Revisiting Noether symmetry of $F(R)$ theory of gravity}
We now understand that to find Noether symmetry for $F(R)$ theory of gravity, it is better to choose the variable, $h_{ij} = a^2 = z$, rather than the scale factor $a$. In this section we show that under such a choice of variable, Noether symmetry of $F(R)$ theory of gravity yields a rather handy conserved current, which can be solved at ease, and it is not required to search for the cyclic co-ordinate.

\be A = B\int\left[F(R)-\lambda\left\{R - 6\left(\frac{\ddot
z}{2z}+\frac{k}{z}\right)\right\}\right]z^{\frac{3}{2}}dt- 2 B\int \left[\sqrt h F_{,R} K \right]d^3 x.\ee
Following the same technique, i.e., varying the action with respect to $R$ and setting it to zero one finds $\lambda = F_{,R}$, which when substituted in (41) and the surface term is accounted for through integration by parts, the action may be expressed in the following non-degenerate and canonical form,
\be A = B\int \Big[ (F - R F_{,R})z^{\frac{3}{2}} -
\frac{3}{2}F_{,R}\frac{\dot z^2}{\sqrt z}-3F_{,RR}\sqrt z \dot
R\dot z + 6kF_{,RR}\sqrt z\Big] dt.\ee
Demanding Noether symmetry one obtains,
\[ \alpha\left[\frac{3}{2}(F - R F_{,R})\sqrt z +
F_{,R}\frac{3\dot
z^2}{4z^{\frac{3}{2}}}-F_{,RR}\frac{3\dot R\dot z}{2\sqrt
z} + 3k\frac{F_{,R}}{\sqrt z}\right]- \beta \left[F_{,RR}\left(R
z^{\frac{3}{2}} + \frac{3\dot z^2}{2\sqrt z} - 6k\sqrt
z\right)+ 3F_{,RRR}\sqrt z \dot R\dot z \right]\]
\be- (\alpha_{,z}\dot z + \alpha_{,R}\dot R)\left(3F_{,R}\frac{\dot
z}{\sqrt z} + 3F_{,RR}\sqrt z \dot R\right)-(\beta_{,z}\dot z +
\beta_{,R}\dot R)(3F_{,RR}\sqrt z \dot z) = 0.\ee
Equating coefficients as usual, following four equations emerge,
\be \frac{\alpha}{2z} - 2\alpha_{,z} =
\frac{F_{,RR}}{F_{,R}}(\beta + 2z \beta_{,z}).\ee
\be F_{,RR}\sqrt z \alpha_{,R} = 0.\ee
\be \beta \frac{F_{,RRR}}{F_{,RR}} + \beta_{,R} +
 \frac{\alpha}{2z}+ \alpha_{,z} + \frac{\alpha_{,R}F_{,R}}{z F_{,RR}}= 0.\ee
\be \alpha\left[\frac{3}{2}(F - R F_{,R})+
3k\frac{F_{,R}}{z}\right]= \beta \left(R
z- 6k\right)F_{,RR}.\ee
For a non linear form of $F(R)$, i.e., $F_{,RR} \ne 0$, equation (45)
implies that $\alpha = \alpha (z)$ only. Thus, equation (44)
demands separation of variable $\beta$ as, $\beta =
\beta_{1}(z)\beta_{2}(R)$, and is expressed as,
\be \frac{\frac{\alpha}{2z} -
2\alpha_{,z}}{\beta_{1}+2z\beta_{1,z}} =
\beta_{2}\frac{F_{,RR}}{F_{,R}} = c_1,\ee
while, equation (46) is expressed as,
\be -\frac{\frac{\alpha}{2z} + \alpha_{,z}}{\beta_{1}} =
\beta_{2}\frac{F_{,RRR}}{F_{,RR}} + \beta_{2,R} = c_2,\ee
where, $c_1$ and $c_2$ are separation constants. It is not difficult to check that middle terms of the above two
equations are the same, i.e., $\beta_{2}\frac{F_{,RR}}{F_{,R}}= \beta_{2}\frac{F_{,RRR}}{F_{,RR}} + \beta_{2,R}$ and so, $c_1 = c_2 = c$. Thus equations (48) and (49) may be solved to obtain,
\be \alpha = m z + n\;\;\;and\;\;\;\beta _1 =
-\frac{1}{2cz}(3mz+n).\ee In view of the solutions (50), equation (47) after simplification takes the following form,
\be 3(m z + n)F - 12 k m F_{,R} - 2 n R F_{,R} = 0,\ee which is
satisfied only under the condition $m = 0$. Hence one can solve equation (51) for $F(R)$ and then equation (48) for $\beta_2(R)$.
The final set of solutions thus obtained is,
\be \alpha = n,\;\;\;\beta = -\frac{n R}{z},\;\;\;F(R) =
R^{\frac{3}{2}},\ee where, $n$ is a constant. The conserved current now reads,
\be {\mathcal F} = \frac{d}{dt}(\sqrt{z R}) = \frac{d}{dt}(a\sqrt{R}),\ee
which is having the same form as obtained earlier in equations (17), (32), (38) and by \cite{cap2008}. Substituting the expression for $R$, the above conserved current
can now be solved directly to obtain the same old solution,
\be a = \left[a_{4}t^4 + a_{3} t^3 + \left(\frac{3
a_{3}^2}{8a_{4}} - k\right)t^2 + a_2 t + a_1\right]^{\frac{1}{2}},\ee
already presented in (33) and (40).

\section{Presence of matter}
The solution (33) or $\big(\; (40)\; or \;(54)\big)$ that we have
presented so far corresponds to vacuum, which may have some
importance in the early curvature dominated Universe, since it
represents transition from decelerating Universe with $a \propto
\sqrt t$ to a power law inflation. To unveil the importance of
$R^{\frac{3}{2}}$ in the late stage of cosmic evolution, one has
to take some form of matter into account. So, let us start again
from the action, \be A = \int \left[B R^{\frac{3}{2}}- M
a^{-3(w+1)} \right]\sqrt{g} d^4 x- 2 B\int \left[\frac{3}{2}\sqrt
{h R}~ K \right]d^3 x,\ee where, $w = \frac{p}{\rho}$ is the state
parameter. Note that, in the radiation dominated era, $w =
\frac{1}{3}$ and so, the second term reads, $M a^{-3(w+1)} =
\frac{\rho_{r0}}{a^4}$ and in the matter era, $w = 0$ and it reads
$M a^{-3(w+1)} = \frac{\rho_{m0}}{a^3}$. Here, $\rho_{r0},\;
and\;\rho_{m0}$ refer to radiation density and the matter density
at the present epoch, which together have been represented by $M$.
Following the same prescription of introducing auxiliary variable
given in (36), and after taking care of the surface term, the
above action can be expressed in the following canonical and
non-degenerate form, \be A = \int\left[-\dot Q\dot z + 2kQ -
\frac{4}{729 B^2} Q^3 - M z^{-\frac{3w}{2}}\right]dt.\ee The field
equations are, \be \ddot Q + \frac{3Mw}{2} z^{-(\frac{3w+2}{2})} =
0,\ee \be \dot Q\dot z + 2k Q -\frac{4}{729 B^2} Q^3 = M
z^{-\frac{3w}{2}},\ee In the radiation dominated era $w =
\frac{1}{3}$ and the field equations admit a solution in the form,
\be \sqrt z = a = b_0 t^{\frac{3}{4}} + b_1,\ee $b_1$ being a
constant of integration, under the condition, $k = 0$ and $B = -
\big(\frac{32 M}{135}\big)$, instead of the standard Friedmann
type, i.e., $a \propto t^{\frac{1}{2}}$. This creates problem in
explaining BBN, since the Universe expands at a much faster rate
than the standard Friedmann model. Further, due to such a high
expansion rate in the radiation era, it becomes difficult for the
seed of perturbations to accumulate matter in the matter dominated
era to form structures. Now, in the matter dominated era $w = 0$
and so is $\ddot Q$. Thus we recover the same solution (33) or
$\big((40)\; or \;(54)\big)$, i.e., \be \sqrt z = a =
\left[a_{4}t^4 + a_{3} t^3 + \left(\frac{3 a_{3}^2}{8a_{4}} -
k\right)t^2 + a_2 t + a_1\right]^{\frac{1}{2}},\ee which is true
for arbitrary curvature parameter, viz., $k = 0,\pm 1$.
\section{Fitting the observational data}

Data fitting of the above unique solution for $R^{\frac{3}{2}}$ in
the matter dominated era, has been presented already by
Capozziello et al \cite{cap2008}. The authors \cite{cap2008} have
chosen a typical unit $H_0 = 1$ and set $H_0t_0 = 1$. Finally
choosing the present value of the deceleration parameter $q_0 =
-0.4$, which corresponds to the present value of the effective
state parameter $w_0 = -0.6$, they reduced the solution to a one
parameter model, for which a nice fit for luminosity - distance
curve with $\Lambda CDM$ is obtained. However, for such parametric
values, the Hubble parameter $H(z)$ versus redshift $z$ curve
(Figure 1) is largely different from the standard $\Lambda$CDM
model at high redshift. This means the solution tracks Friedmann
like matter dominated era $q = 0.5$ only transiently, which is
also apparent from the deceleration parameter $q$ versus redshift
$z$ curve (fig. 2). Therefore it is not clear how the authors
claimed \cite{cap2010} that the matter dominated era in their
model remains unaltered. Further, choosing the present value of
deceleration parameter $q_0 = -0.4$ a-priori, is unphysical, since
WMAP - 7 year data suggests $w_0  = - 1.1 \pm 0.14$, which
corresponds to $-1.36 \le q_0 \le -0.94$, albeit it is model
dependent. For the required fit, the authors \cite{cap2008} also
took some finite value of the curvature parameter $k$, viz., $k =
-0.49$. Here we observe that the only power law solution that
satisfies the field equation in the radiation dominated era,
requires $k = 0$. So, we are required to set $k = 0$, in the
matter dominated era also. Therefore, to understand the model in
some detail, we present yet another fit keeping $q_0$ arbitrary.
Note that one can set $a_1 = 1$ in solution (60) without any loss
of generality. Still, we are left with three parameters, viz.,
$a_2,~a_3$ and $a_4$. We have studied following two cases, keeping
the presently accepted value $H_0 t_0 = 1$ and $h = 0.72$ in both.
\begin{figure}
[ptb]
\begin{center}
\includegraphics[
height=2.034in, width=2.8in]
{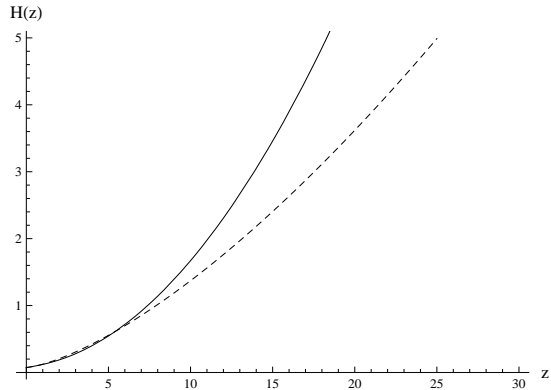} \caption{
Hubble parameter $H(z)$ versus the Redshift parameter $z$ for the present model \cite{cap2008} (solid curve) shows considerable deviation at large
redshift, with the standard $\Lambda$CDM model (dashed curve).}
\end{center}
\end{figure}

\begin{figure}
[ptb]
\begin{center}
\includegraphics[
height=2.034in, width=2.8in]
{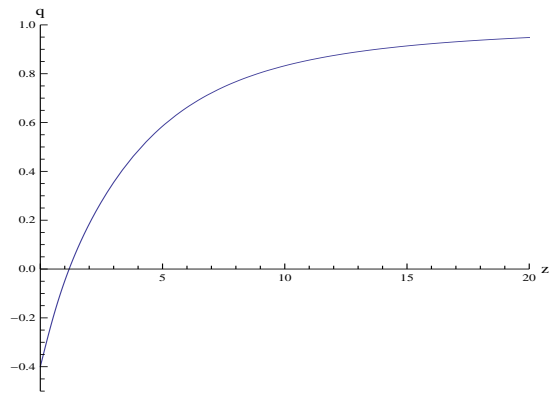} \caption{
Deceleration parameter $q$ versus the Redshift parameter $z$ for the present model \cite{cap2008} shows that Friedmann like matter dominated era
$q = 0.5$ is only transient. This fact tells upon the structure formation.}
\end{center}
\end{figure}

\subsection{$H_0 t_0 = 1$.}
Fixing up $H_0 t_0 = 1$, one can now express, one of the
parameters of the theory (we have chosen $a_2$) in terms of the
other two (viz., $a_3$ and $a_4$ here). The important observation
is that the fit with SNIa data is almost perfect for a truly wide
range of parametric values of $a_3$ and $a_4$. We briefly present
our result in table 1. The table depicts that the present value of
the state parameter $w_{e0}$, remains close to $-1$ for a huge
range of parametric values of $10^{-3} <a_4 < 10^6$ ($a_4 \ne 0$)
and $10^{-5} < a_3 < 10^6$. Within such a range of values the
transition redshift $z_a$ always remains close to $1$. The
Redshift-Distance modulus curve fits almost perfectly with the
$\Lambda CDM$ model and thus the SNIa-data in disguise, taking the
density parameter of the baryonic and nonbaryonic dark matter
$\Omega_m = 0.26$ and the rest, $ \Omega_{\Lambda} = 0.74$, as the
density parameter of the dark energy. One of such fits has been
shown in Fig. 3. Despite such amazingly attractive result, some
serious problems have also been encountered. Almost in all the
cases the redshift value is $z \approx 200$ at around $t = 3.75
\times 10^{-4}$ Billion years. Note that according to the standard
model it is the time after Big-Bang, when photons decoupled and
whose redshift value is $z_{dec} \approx 1100$, as confirmed by
WMAP data. This is practically a huge problem. Further,
considerably large deviation from the standard model in the early
matter dominated epoch is clearly visible from Fig. 4, where a
plot of the Hubble parameter $H(z)$ versus the redshift parameter
$z$ is drawn with the parametric values $a_4 = 1, a_3 = 0.1$. The
two, deviates even further at a large redshift and the feature is
the same for all other parametric values. Such a deviation from
the standard model in the early matter dominated era is
inconsistent with the WMAP data and LSS \cite{amen2007}. Finally,
Fig. 5 is the effective state parameter $w_{e}$ versus redshift
$z$ plot for the same above parametric values. It shows that the
effective state parameter $w_e$ remains close to $0.33$ at large
$z$, which clearly depicts that the early matter dominated era is
far from the standard Friedmann model ($w_e = 0$) and tracks
$a\propto t^{\frac{2}{3}}$ only transiently.

\begin{figure}
[ptb]
\begin{center}
\includegraphics[
height=2.034in, width=2.8in] {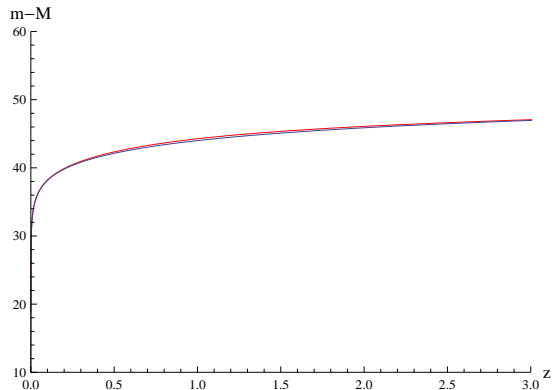} \caption{Distance modulus $(m-M)$ versus redshift $z$ plot of the present model (blue/lower), shows perfect
fit with the $\Lambda$CDM model (red/upper) taking $a_4 = 1$ and $a_3 = 0.1$. The same feature is observed almost for all parametric values of $a_3$
and $a_4$ (colour figure online).}
\end{center}
\end{figure}

\begin{figure}
[ptb]
\begin{center}
\includegraphics[
height=2.034in, width=2.8in] {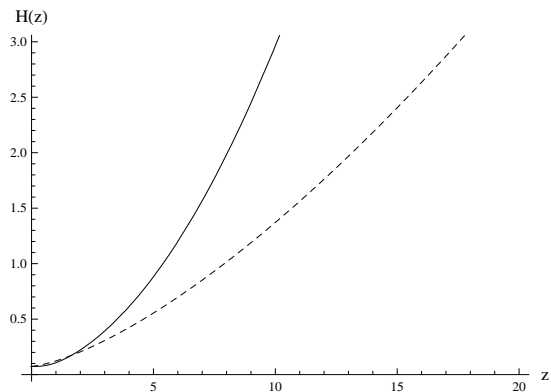} \caption{
Hubble parameter $H(z)$ versus the Redshift parameter $z$ for the present model (solid curve) setting $H_0t_0 = 1, h = 0.72$, $a_4 = 1$ and $a_3 = 0.1$ still shows a wide deviation at large redshift, with the standard $\Lambda$CDM model (dashed curve).}
\end{center}
\end{figure}

\begin{figure}
[ptb]
\begin{center}
\includegraphics[
height=2.034in, width=2.5in] {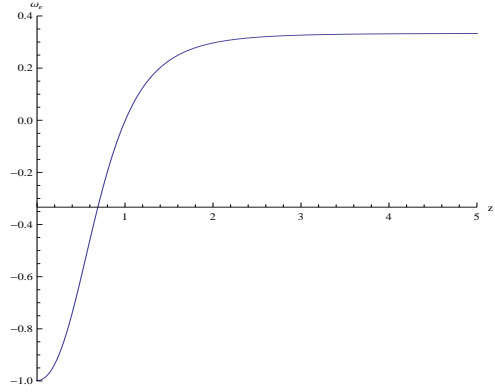} \caption{The effective state parameter $w_{e}$ versus the Redshift parameter $z$ for the present model
($a_4 = 1, a_3 = 0.1$) shows that in the early epoch of matter domination $w_e = 0.33$, which indicates that the scale factor during the matter
phase grows as $t^{\frac{1}{2}}$, instead of the standard $t^\frac{2}{3}$ - law. This is grossly inconsistent with cosmological observations viz.,
WMAP and LSS.}
\end{center}
\end{figure}

\begin{table}
\centering
\caption{Data fitting with $H_0 t_0 = 1$ and $h = 0.72$}\label{*1}
\begin{tabular}{| c | c | c | c |}\hline
  % after \\: \hline or \cline{col1-col2} \cline{col3-col4} ...\\
$a_4$ & $a_3$ & $w_{e0}$ & $z_a$  \\\hline
$10^5$ & $10^4$ & -$0.998$ & $0.72$  \\\hline
$10^5$ & $10^5$ & $-0.984$ & $0.70$ \\\hline
$10^5$ & $10^6$ & $-0.866$ & $0.68$  \\\hline
$10$ & $0$ & $-1$ & $1$\\\hline
$1$ & $0.1$ & $-1$ & $0.68$\\\hline
$1$ & $0.001$ & $-0.999$& $1$\\\hline
$5$ & $1$ & $-0.997$ & $1$\\\hline
$0.01$ & $0.00079$ & $-1$ & $1$ \\\hline
$10^{-3}$ & $7.78 \times 10^{-5}$ & $-1$ & $0.71$ \\\hline

\end{tabular}
\end{table}

\subsection{$H_0 t_0 = 1$ ~and~ $z_{dec} = 1100$, for $t = 3.75 \times 10^5$ years.}

In the preceding section we have noticed that the unique solution
obtained for $R^{\frac{3}{2}}$ action in the matter dominated era
shows excellent fit with $\Lambda CDM$ model and hence SNIa data
in disguise, for a huge range of parametric values, but the
redshift value is nowhere near that at decoupling. Hence, here we
choose WMAP data  $(z_{dec})$ a-priori, which fixes up yet another
parameter of the theory, viz., $a_3$, in the present case. Thus,
we end up with a one-parameter model. Here, we observe that again
for a wide range of parameter, $a_4 > 10 ^{-5}$, with $a_4 \ne 0$,
the Redshift-Distance modulus curve fit is almost perfect, like
the one presented in Fig. 2. The present value of the effective
state parameter is $w_{e0} \approx -0.35$ in almost all the cases,
which is nowhere near $\Lambda CDM$ model. But then, $w_{e0}$ is
model dependent, and so it creates no problem as such. However,
the transition redshift is unacceptably large, the minimum of
which is $z_{a} = 4.2$. Further, the Hubble parameter $H(z)$
versus redshift $z$ plot (Fig. 6), here again shows huge deviation
which increases rapidly with $z$. The only difference is the
Hubble parameter for the present model here, remains far below
than that corresponding to the standard model. The effective state
parameter again tracks $0.33$, instead of $w_e = 0$, in the early
matter dominated epoch, which is grossly inconsistent with the
LSS.

\begin{figure}
[ptb]
\begin{center}
\includegraphics[
height=2.034in, width=2.8in] {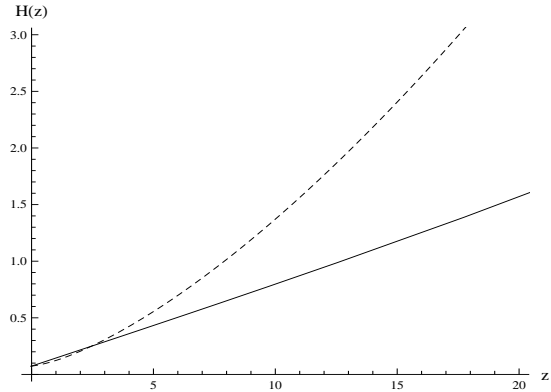} \caption{Hubble parameter $H(z)$ versus the Redshift parameter $z$ for the present model (solid curve)
choosing for $H_0 t_0 = 1, a_4 = 1$ (note that here $a_3$ has been fixed from WMAP data) again shows a wide deviation with the standard
$\Lambda$CDM model (dashed curve). }
\end{center}
\end{figure}

\section{Palatini Variational formalism}
Since Noether symmetry approach via metric formalism does not
yield anything other than $F(R) \propto R^{\frac{3}{2}}$, which
runs into problem as discussed, so it is viable to attempt
palatini variational technique for the same. In this technique,
the connection (usually torsion free) $\Gamma^\alpha_{\mu\nu}$,
out of which the Ricci tensor is defined, is treated to be
independent of the space-time metric, i.e., it is no longer the
Levi-Civita connection. As a result, the action is varied both
with respect to the metric tensor and the connection. In the case
of Einstein-Hilbert action, Palatini variational techniques
reproduces general theory of relativity (GTR) making the
connection to be the Levi-Civita one. The only difference is that
the surface term does not arise in the process and thus the
equivalence of the surface term with the black hole entropy is
lost. However, for extended theory of gravity the two variational
techniques yield different field equations and hence
different Physics altogether. These issues have been thoroughly discussed in a recent review \cite{cap11}.\\
To cast the action containing $F(R)$ gravity in the Canonical form in view of Palatini variational formalism it is required to first form a
scalar-tensor equivalence of the theory. Starting from the action

\be S(\Gamma^\alpha_{\mu\nu}, g_{\mu\nu}) = \int \sqrt{-g} d^4 x F(\mathcal{R}) + S_m,\ee
where, $\mathcal{R} = \mathcal{R}(\Gamma^\alpha_{\mu\nu}, g_{\mu\nu})$ and $S_m$ is the matter action which is usually assumed not to be a function of the connection. Under conformal transformation the above action leads to the following special type of
Brans-Dicke action with $\omega = -\frac{3}{2}$,

\be S = \int  \sqrt{-g} d^4 x [\phi R + \frac{3}{2 \phi} \phi_{,\mu}\phi^{,\mu} - V(\phi)] + S_m,\ee
where,
\be \mathcal{R} = R + \frac{3}{2 \phi} \phi_{,\mu}\phi^{,\mu} - \frac{3}{2}\Box\phi;\;\; V(\phi) = \phi\chi(\phi) - F[\chi(\phi)];\;\; \phi
= F'(\chi), \chi = \mathcal{R}.\ee
It has been claimed \cite{ros08} that Noether symmetry for the action (62) in the spatially flat Robertson-Walker metric in vacuum or being
associated with pressureless dust leads to $F(R) \propto R^n$, along with a conserved current which is related to the Newton's gravitational
constant and the present matter content of the Universe. Thus Palatini variational technique apparently yields a better result. However, this
is again of no practical use. The reason is, Noether symmetry in this technique admits only a single curvature invariant term corresponding
to a particular choice of $n$, while, a renormalizable, asymptotically free and unitary theory of gravity requires curvature squared terms
along with the linear one in the early Universe and, as already demonstrated, in the late stage of cosmic evolution presently available cosmological data can not again be explained in view of a single curvature invariant term. Hence, we conclude that Noether symmetry of $F(R)$ theory of gravity does not help in any sense. Situation might improve if the configuration space is enlarged by including a scalar field in addition, forming a $K(\phi, R)$ theory \cite{cap11} or if Palatini variational technique admits Noether symmetry for an action containing $R^m + R^n$, with $m \ne n$. These we pose in future communications.

\section{Summary}
We have made a detailed analysis corresponding to the cosmological evolution with $F(R) = R^{\frac{3}{2}}$ theory of gravity, in the Robertson-Walker
minisuperspace model. We enlist the present findings.\\
1. Noether symmetry of $F(R)$ theory of gravity does not admit anything other than $F(R) = R^{\frac{3}{2}}$ in the Robertson-Walker
minisuperspace model.\\
2. The speciality of such a curvature invariant term is apparent under a change of variable from the scale factor $a$ to the basic variable
$h_{ij} = a^2$, since $h_{ij}$ becomes cyclic. Thus we conclude that Noether symmetry is in-built in $R^{\frac{3}{2}}$ theory of gravity,
for Robertson-Walker minisuperspace model. Under the change of variable, the field equation looks pretty simple and is solved at once.\\
3. If one tries to find Noether symmetry of $F(R)$ theory of gravity, using the basic variable $h_{ij}$, the conserved current is solved
directly, and it is no more required to find cyclic co-ordinate.\\
4. Since Noether symmetry of  $F(R)$ theory of gravity is independent of the configuration space variables, so it is an in-built symmetry. \\
5. The solution in the radiation era ($a \propto
t^{\frac{3}{4}}$), is substantially different from the standard
Friedmann solution $a \propto t^{\frac{1}{2}}$, which indicates
much faster ($1.5$ times) expansion rate of the Universe. This
clearly puts up severe problem in explaining Nucleosynthesys.\\
6. The solution in the early matter dominated era tracks
$t^{\frac{1}{2}}$ power law, rather than standard
$t^{\frac{2}{3}}$, like any $R^{-n}$, with $n > 0$ model, which again
creates problem in structure formation and fitting WMAP data, as discussed
earlier \cite{amen2007}.\\
7. Since Palatini variational technique does show symmetry for
$F(R) \propto R^n$ in vacuum and in the presence of pressure-less
dust, it allows only a single curvature invariant term. Both the
early Universe and the late one require more than one curvature
invariant term and so Palatini variational technique is also of no
use in this context.\\
Thus, although it has been demonstrated \cite{cap2010} that
$R^{\frac{3}{2}}$ might be useful to explain some of the presently
available data and perhaps solve the MOND problem at small scales
\cite{men2011}, still unless supplemented by Einstein-Hilbert
term, $R^{\frac{3}{2}}$ term alone suffers from serious disease.
Since, Noether symmetry of $F(R)$ theory of gravity does not admit
anything other than $R^{\frac{3}{2}}$, in the Robertson-Walker
minisuperspace model so, either we have to abandon Noether
symmetry or try to involve more degrees of freedom in the theory,
by incorporating either scalar field or working with some
anisotropic minisuperspace model. This we pose in a forthcoming article \cite{kaushik12}.\\

\end{document}